\documentclass[5p,times]{elsarticle}
\usepackage{epsfig}
\usepackage{graphicx}
\usepackage{amssymb}
\usepackage[figuresright]{rotating}
\usepackage{amssymb}
\def\bea {\begin{eqnarray}}
\def\eea {\end{eqnarray}}

\def\be {\begin{equation}}
\def\ee {\end{equation}}

\usepackage{float}
\usepackage{framed}
\usepackage{amsmath}
\usepackage{subfigure}
\usepackage{textcomp}
\usepackage{multirow}
\usepackage{enumerate}
\usepackage{hyperref}
\biboptions{numbers,sort&compress}
\begin{document}
\begin{frontmatter}

\title{ Cascade and Omega productions from Pb+Pb Collisions at LHC energy}

\author{\it Purabi Ghosh$^1$, Jajati K. Nayak$^{*2}$, Sushant K. Singh$^2$ and Santosh K. Agarwalla$^1$}
\ead{$^*$jajati-quark@vecc.gov.in}

\address{$^1$ Dept. of Applied Physics and Ballistics, F. M. University, Balasore, Odisha. \\
$^2$ Variable Energy Cyclotron 
Centre, 1/AF Bidhan Nagar, Kolkata-700064, India.}            

\begin{abstract}
Production of multi strange hadrons like cascade ($\Xi$) and omega($\Omega$) baryons are studied microscopically using rate equation at $\sqrt{s_{NN}}$=2.76 TeV, Large Hadron Collider(LHC) energy. The rate equations for $\Xi$, $\Omega$ are solved simultaneously with other strange hadrons in an expanding medium. The results for $\Xi$ and $\Omega$ are compared with the data obtained from Pb-Pb collisions at $\sqrt{s_{NN}}$= 2.76 TeV from ALICE experiments. The ratio of yields of $\Xi$ and $\Omega$ to $\pi$ are analysed with various initial conditions and compared with the experimental observations made for various charge particle multiplicities.   

\end{abstract}

\end{frontmatter}
\section{\label{sec:intro} Introduction}
Recent measurements of multi strange baryons, $\Xi$ and $\Omega$ from p-p \& p-Pb and Pb-Pb collisions at LHC energies \cite{alicenature17,gyula_alice,multistrange_alice_plb14} show interesting results that led to several intense theoretical activities. The ratios of the yield of $\Xi$ and $\Omega$ baryons to pions are observed to be enhanced with multiplicity at $\sqrt{s_{NN}}$=2.76 TeV, Pb-Pb collisions\cite{multistrange_alice_plb14}. A similar trend has been observed recently in high multiplicity p-p and p-Pb collisions at $\sqrt{s_{NN}}$= 7 TeV and 5.02 TeV respectively\cite{alicenature17,gyula_alice,adam16}. From the measurements of Pb-Pb collisions at 2.76 TeV, data of $\Xi/\pi$ and $\Omega/\pi$ are not available at lower multiplicities, below $dN_{ch}/d\eta$=35 (corresponding to 60-80$\%$ centrality with $N_{\text{part}}$=22.5)~\cite{multistrange_alice_plb14} to compare with the measurements from p-p (7 TeV) and p-Pb (5.02 TeV) collisions\cite{alicenature17,gyula_alice,adam16}. However, when all available data of $\Xi /\pi$ and $\Omega/\pi$ at various multiplicities of different colliding energies are put together, it indicates that there is a steady rise in the multi strange hadron yield with multiplicity across all collision systems and then there might be a saturation. But this trend of smooth rise is not manifested strongly in case of $\Xi/\pi$ and that is clear from the data point corresponding to the lowest multiplicity of 2.76 TeV Pb-Pb collisions\cite{multistrange_alice_plb14,alicenature17}. These data are not explained till now with microscopic detail. Here the attempt has been made to understand the microscopic productions of multi strange hadrons from Pb-Pb collisions at $\sqrt{s_{NN}}$= 2.76 TeV. 

Strange meson productions in hadronic phase have been studied using several models~\cite{rafelski82,kapusta86,BT07,
CORS,MG04,andronic06,jknacta06,jknprc10,tawfik09}. However none of these models explain the production and evolution of multi strange baryons and their enhancement over p-p collisions. Statistical hadronisation model evaluated the integrated yield at those energies assuming common chemical freeze out temperature for all species\cite{andronic06,andronicplb09} including RHIC and LHC energies. However it could not explain the ratios of multi-strange hadrons at 0-20\% centrality of Pb-Pb collision at $\sqrt{s_{NN}}$= 2.76 TeV LHC energy  while fitting with $p/\pi$ ratio. 
Similarly productions of kaons and anti kaons at higher colliding energies such as at RHIC and LHC (also at higher SPS energies) have been explained using models with strange quark evolution assuming a QGP phase\cite{BT07,jknprc10}. But the multi-strange productions are not explained there. 

Enhancement of multi strange baryons at SPS energy has been attempted using URQMD in \cite{bassplb99} but the data were not explained well and it was argued that the enhanced productions might be due to the topological defects arising from the formation of disoriented chiral condensates (DCC) at the initial stages of collisions where the density is too high. Authors in \cite{kolomeitsev12} have made a novel attempt to explain HADES data using minimal statistical hadronisation model and tried to explain the ratio $\Xi^-/\Lambda$ and $\Omega^-/\Xi^-$ in \cite{kolomeitsev15} but could not reproduce the data although got the same trend.

In this article, we focus on the microscopic production of multi-strange baryons $\Xi$ \& $\Omega$ and their evolution for the first time in an expanding hot-dense system as produced in relativistic heavy ion collisions using rate equation. Extensive analysis has been done with various initial conditions and the results are compared with the observations of $\Xi/\pi$ and $\Omega/\pi$ from Pb-Pb collisions at $\sqrt{s_{NN}}$=2.76 TeV\cite{multistrange_alice_plb14}. 

In the next section the production of $\Xi$ and $\Omega$  and their interactions in hadronic matter is discussed in detail. In section \ref{sec:strange_evolve}, the evolution of strange hadrons are discussed along with secondary productions using rate equations. In this section, the equations for temperature and baryon chemical potential evolution are also highlighted. Then the results are presented in section \ref{sec:results} and compared with the experimental observation. Finally, the work has been summarised in section \ref{sec:summary}.
\section{\label{sec:multistrangeprod} Multi strange productions in hadronic matter}
When the energy deposition in heavy ion collision is more than certain threshold value, an initial quark gluon system may be produced. The hadrons then are produced from the quarks due to hadronization as system expands. 
On the other hand when the energy deposition is less, then an initial hadronic state is plausible. Then the hadronic system evolves with secondary collisions and continue till the freeze out of hadronic species occur. The produced system encounters a hadronic medium whatever may be the energy deposition such as at RHIC and LHC. In the present article the study focuses on the production and evolution of multi-strange hadrons $\Xi$ and $\Omega$ in a hadronic medium. To study cascade and omega it is important to discuss the hadronic interactions those govern the system. 

The interactions considered for $\Xi$ and $\Omega$ productions are as follows;  
$ \bar{K}N \rightarrow K \Xi$, 
$ \bar{K}\Lambda \rightarrow \pi \Xi $,
$ \bar{K}\Sigma \rightarrow \pi \Xi $,
$ \Lambda\Lambda \rightarrow N \Xi$,
$ \Lambda\Sigma \rightarrow N \Xi $,
$ \Sigma\Sigma \rightarrow N \Xi$,
$\Lambda \bar{K}\rightarrow \Omega^{-} K^0$,
$\Sigma^{0} \bar{K}\rightarrow \Omega^{-} K^0$,
$\bar{p} p\rightarrow \Omega^{-} \Omega$,
$ p\bar{p} \rightarrow \Xi \bar{\Xi}$,
$ \pi\Xi \rightarrow \Omega K$, where $N$ represents nucleon. We have also considered productions of other strange mesons and baryons along with $\Xi$ and $\Omega$ which are discussed in the next section. Isospin combinations are also taken into account. The production of the strange hadrons is then studied using transport equation which is discussed later.

Along with these channels the inverse processes have also been considered using principle of detailed balance~\cite{cugnon84}. There are also other $2\rightarrow 3$ and $2\rightarrow 4$ channels those contribute to the strange productions, but their rates of production are much less due to phase space factor, hence not considered here. All hadronic interactions for strange productions are broadly categorised as meson-meson, meson-baryon and baryon-baryon interactions. Each category has dominance over the other in different domain of colliding energies or depending on the system with mesonic or baryonic abundances.
The channels for single strange(s=-1)productions and their cross sections are  in ~\cite{Brown1,amslar08}. See for details~\cite{jkn19}.
\subsection{\bf Cross sections of Cascade($\Xi$) and Omega ($\Omega$) production}
The strangeness content in $\Xi$ (S=-2) and $\Omega$ (S=-3) is more. Hence the production of these baryons are mostly from strangeness exchange reaction. To produce a baryon of S=-2 or -3 is more expensive and less probable from the reactions involving non strange hadrons in the initial channels. 
\subsubsection{\bf Cross sections for $\Xi$ production}
The channels involved in cascade productions are $\Lambda\Lambda\rightarrow N\Xi$, $\Lambda\Sigma\rightarrow N\Xi$, $\Sigma\Sigma\rightarrow N\Xi$, $\bar{K} \Lambda \rightarrow \pi \Xi$, $\bar{K} \Sigma \rightarrow \pi \Xi$, $\bar{K} N \rightarrow K \Xi$, $\bar{p} p \rightarrow \Xi \bar{\Xi}$, $K \Omega \rightarrow \pi \Xi$. Some of them are strangness exchange reactions and some are not.

The cross sections from strangeness exchange reactions have been considered from a gauged flavor SU(3) symmetry Langragian density~ \cite{liprc85,linpa02} as follows,
\begin{align}
 \mathcal{L} &= i\text{Tr}\left(\bar{B}\not\partial B\right)+\text{Tr}\left[\partial_{\mu} P^{+}\partial^{\mu}P\right] \nonumber \\
 &+ g' \text{Tr}\left[ \left(2\alpha-1\right)\bar{B}\gamma^5\gamma^{\mu} B\partial_{\mu}P+ \bar{B}\gamma^5\gamma^{\mu}\left(\partial_{\mu} P\right)B\right]
 \label{lagrangianeq1}
\end{align}
 where, 
 \[
 B=
\begin{bmatrix}
     \frac{\Sigma^0}{\sqrt{2}}+\frac{\Lambda}{\sqrt{6}}& \Sigma^+ & p \\
      \Sigma^{-} & \frac{-\Sigma^{0}}{\sqrt{2}}+\frac{\Lambda}{\sqrt{6}} & n \\ 
      -\Xi^{-} & \Xi^{0} & -\sqrt{\frac{2}{3}}\Lambda\\
\end{bmatrix}
\]
 \[
 P=\frac{1}{\sqrt{2}}
  \begin{bmatrix}
     \frac{\pi^{0}}{\sqrt{2}}+\frac{\eta_8}{\sqrt{6}}+\frac{\eta_1}{\sqrt{3}} & \pi^{+} & K^{+}\\
     \pi^{-} & \frac{-\pi^0}{\sqrt{2}}+\frac{\eta_8}{\sqrt{6}}+\frac{\eta_1}{\sqrt{3}} & K^0\\
     K^{-} & \bar{K}^{0} & -\sqrt{\frac{2}{3}}\eta_8+\frac{\eta_1}{\sqrt{3}}\\
  \end{bmatrix}
\]
with $B$ and $P$ representing baryon and pseudoscalar meson octects respectively. $P$ is the linear combination of both pseudoscalar octet ($\pi, K, \eta_8$) and singlet ($\eta_1$) mesons. $g'$ is the universal coupling constant between baryons ($B$) and pseudoscalar mesons ($P$). $\alpha$ is a parameter obtained from the coupling constants of $D$-type and $F$-type interactions of $P$ and $B$ and value is taken to be 0.64 ~\cite{adelseck90}. 

To consider the vector mesons in the interactions of baryons and pseudo scalar mesons, vector mesons are treated as gauge particles and taken care by replacing partial derivative $\partial_{\mu}$ with covariant $D_{\mu}$, where
\begin{equation}
 D_{\mu}=\partial_{\mu}-ig[V_{\mu}]\label{covderiv}.
\end{equation}
and $g$ is the other universal coupling constant that tells about the strength of vector meson interaction with pseudo scalar mesons and baryons~\cite{linpa02}. For details refer \cite{jkn19}. Assuming $SU(3)$ invariant tensor interactions of $D$ and $F$ type, the interaction lagrangian then can be written as, 
\begin{equation}
 \mathcal{L}^t =\frac{g^t}{2m} \text{Tr}[(2\alpha-1)\bar{B}\sigma^{\mu\nu}B\partial_{\mu}V_{\nu}+\bar{B}\sigma^{\mu\nu}(\partial_\mu V_\nu)B] 
\end{equation}
where $g^t$ is the universal tensor coupling constants and obtained from the empirical values of coupling of $\rho-N$ tensor interactions given by $g^t_{\rho NN}=19.8$ ~\cite{holzenkamp89}. $m$ represents the degenerate baryon mass. It may be noted that the contributions from the axial vector mesons $a_1(1260)$ and $K_1(1270)$ are expected to be small because of their masses and hence not considered in this work.

The cross sections for the strangeness exchange reactions $\bar{K} \Lambda \rightarrow \pi \Xi$, $\bar{K} \Sigma \rightarrow \pi \Xi$ have been calculated considering the amplitude in the Born approximations using coupled channel approach~\cite{linpa02}. Finite size effect of the hadrons at the interaction vertices have been taken care by considering monopole form factor.  

The cross sections are given by \cite{chen04} 
\begin{eqnarray}
\sigma_{\bar{K}\Lambda\rightarrow \pi \Xi} &=& \frac{1}{4}\frac{p_{\pi}}{p_{\bar{K}}}\mid M_{\bar{K}\Lambda\rightarrow \pi \Xi}\mid^{2} \nonumber\\
\sigma_{\bar{K}\Sigma\rightarrow \pi \Xi} &=& \frac{1}{12}\frac{p_{\pi}}{p_{\bar{K}}}\mid M_{\bar{K}\Sigma\rightarrow \pi \Xi}\mid^{2}
\label{casccross1}
\end{eqnarray}
where,
$\mid M_{\bar{K}\Lambda\rightarrow \pi \Xi}\mid^{2}=34.7~\frac{s_0}{s}$ and $\mid M_{\bar{K}\Sigma\rightarrow \pi \Xi}\mid^{2}=318(1-\frac{s_0}{s})^{0.6}\times(\frac{s_0}{s})^{1.7}$~with $p_i$ denoting the centre of mass momentum and $s_0=\sum_i{m_i}$ is the threshold energy and $m_i$ are masses of incoming particles. The inverse reactions from the principle of detailed balance are obtained as 
\begin{eqnarray}
\sigma_{\pi \Xi\rightarrow \bar{K}\Lambda } &=& \frac{1}{3}\frac{p_{\bar{K}}^2}{p_{\pi}^2} \sigma_{\bar{K}\Lambda\rightarrow \pi \Xi}\nonumber\\
\sigma_{\pi \Xi\rightarrow \bar{K}\Sigma} &=& \frac{p_{\bar{K}}^2}{p_{\pi}^2} \sigma_{\bar{K}\Sigma\rightarrow \pi \Xi}
\label{casccross1in}
\end{eqnarray}
Similarly, the cross sections for $Y Y \rightarrow N Y$: $\Lambda\Lambda\rightarrow N\Xi$, $\Lambda\Sigma\rightarrow N\Xi$, $\Sigma\Sigma\rightarrow N\Xi$ are as follows;
\begin{eqnarray}
 \sigma_{\Lambda\Lambda\rightarrow N\Xi}&=&37.15\frac{p_N}{p_\Lambda}\left(\sqrt{s}-\sqrt{s_0}\right)^{-0.16} ~~\text{mb} \nonumber\\
 \sigma_{\Lambda\Sigma\rightarrow N\Xi}&=&25.12\left(\sqrt{s}-\sqrt{s_0}\right)^{-0.42} ~~\text{mb}\nonumber\\
 \sigma_{\Sigma\Sigma\rightarrow N\Xi}&=&8.51\left(\sqrt{s}-\sqrt{s_0}\right)^{-0.395} ~~\text{mb}
 \label{casccross2a}
\end{eqnarray}
Above parametrisation is valid for $0<(\sqrt{s}-\sqrt{s_0})<0.6$ GeV, which is allowed for our calculation. The calculations are only for the Born approximation. 
The other category of reactions producing cascade is $\bar{K} B\rightarrow K \Xi$ or $\bar{K} N\rightarrow K \Xi$. Their cross sections were measured experimentally \cite{bellefon72}
and recently compared with a phenomenological calculation in ~\cite{sharov11}. Following parameterised cross sections of isospin channels have been used. 
\begin{eqnarray}
\sigma_{K^-p \rightarrow K^+\Xi^-}&=&235.6\left(1-\frac{\sqrt{s_0}}{\sqrt{s}}\right)^{2.4}\left(\frac{\sqrt{s_0}}{\sqrt{s}}\right)^{16.6} \text{ mb} \nonumber\\
\sigma_{K^-p \rightarrow K^0\Xi^0}&=&7739.9\left(1-\frac{\sqrt{s_0}}{\sqrt{s}}\right)^{3.8}\left(\frac{\sqrt{s_0}}{\sqrt{s}}\right)^{26.5} \text{ mb}\nonumber\\
\sigma_{K^-n \rightarrow K^0\Xi^-}&=&235.6\left(1-\frac{\sqrt{s_0}}{\sqrt{s}}\right)^{2.4}\left(\frac{\sqrt{s_0}}{\sqrt{s}}\right)^{16.6} \text{ mb}.
\label{casccross3}
\end{eqnarray}
Averaging over isopin channels one can have the cross section for $\bar{K}N\rightarrow K\Xi$ channel as
\begin{eqnarray}
 \sigma_{\bar{K}N->K\Xi}=0.5\left[\sigma_{K^-p \rightarrow K^+\Xi^-}+\sigma_{K^-p \rightarrow K^0\Xi^0}+\sigma_{K^-n \rightarrow K^0\Xi^-}\right]
\end{eqnarray}
The parametrisation here is valid within $0 \leq \left({\sqrt{s}-\sqrt{s_0}}\right) \leq 1 ({\text {GeV}})$. 

The $\Xi$ productions from other important category of reaction where the initial channel doesn't contain any strange hadron is $B \bar{B} \rightarrow \Xi \overline{\Xi}$ \emph{i.e.} $p \bar{p} \rightarrow \Xi^- \overline{\Xi}^+$ and $p\bar{p}\rightarrow \overline{\Xi}^0\Xi^0$. The cross section has been calculated using quark gluon string model\\ (QGSM) and compared with experiment \cite{kaidalov94}. The cross sections of two isospin (outgoing) channels are related as follows, 
$\sigma_{\bar{p}p \rightarrow \overline{\Xi}^+\Xi^-}=16\sigma_{\bar{p}p\rightarrow \overline{\Xi}^0\Xi^0}$
where,
\begin{equation}
 \sigma_{\bar{p}p \rightarrow \overline{\Xi}^0\Xi^0}=\frac{16}{81\pi}\frac{[\sigma_{\bar{p}p\rightarrow \bar{\Lambda}\Lambda}]^2}{2\Lambda_1}\text{exp}\left[\Lambda_1 t_{DC}\right]
\label{crosslambdapp}
\end{equation}
Here the cross section of $\Xi$ production is related to the cross section of $\bar{p} p\rightarrow \bar{\Lambda}\Lambda$. The parameter $\Lambda_1$ appearing in the Eq.~\ref{crosslambdapp} is the slope of the differential cross section of $\bar{p}p\rightarrow \bar{\Lambda}\Lambda$ and the value is taken to be 9 $\text{GeV}^{-2}$ ~\cite{kaidalov94}. For the other factor appearing in the exponent please see \cite{kaidalov94,jkn19}. 
\begin{figure}[t]
\begin{center}
\includegraphics[scale=0.3]{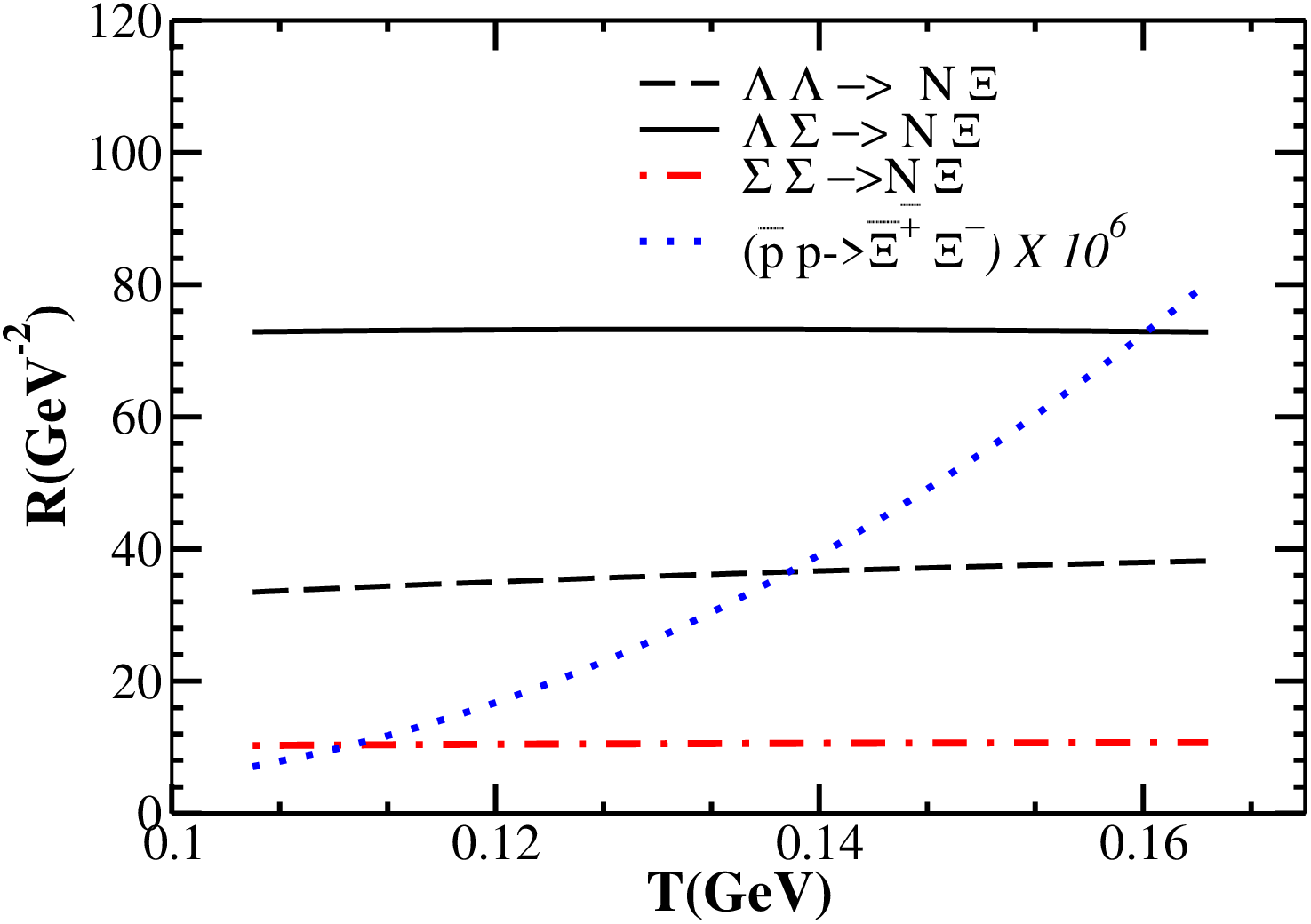}
   \caption{Rate (R=$\langle \sigma v\rangle$)of cascade production with temperature from reactions $YY \rightarrow N\Xi$ and $pp\rightarrow \Xi \bar{\Xi}$. Dashed,solid, dot-dashed(color-red onlie) and dotted (color online-blue)lines represent the contributions from $\Lambda \Lambda\rightarrow N\Xi$, $\Lambda \Sigma\rightarrow N\Xi$ and $\Sigma \Sigma\rightarrow N\Xi$, $p\bar{p}\rightarrow \Xi\bar{\Xi}$  respectively.}
\label{fig_cascaderate1}
\end{center}
\end{figure}
\begin{figure}[t]
\begin{center}
\includegraphics[scale=0.3]{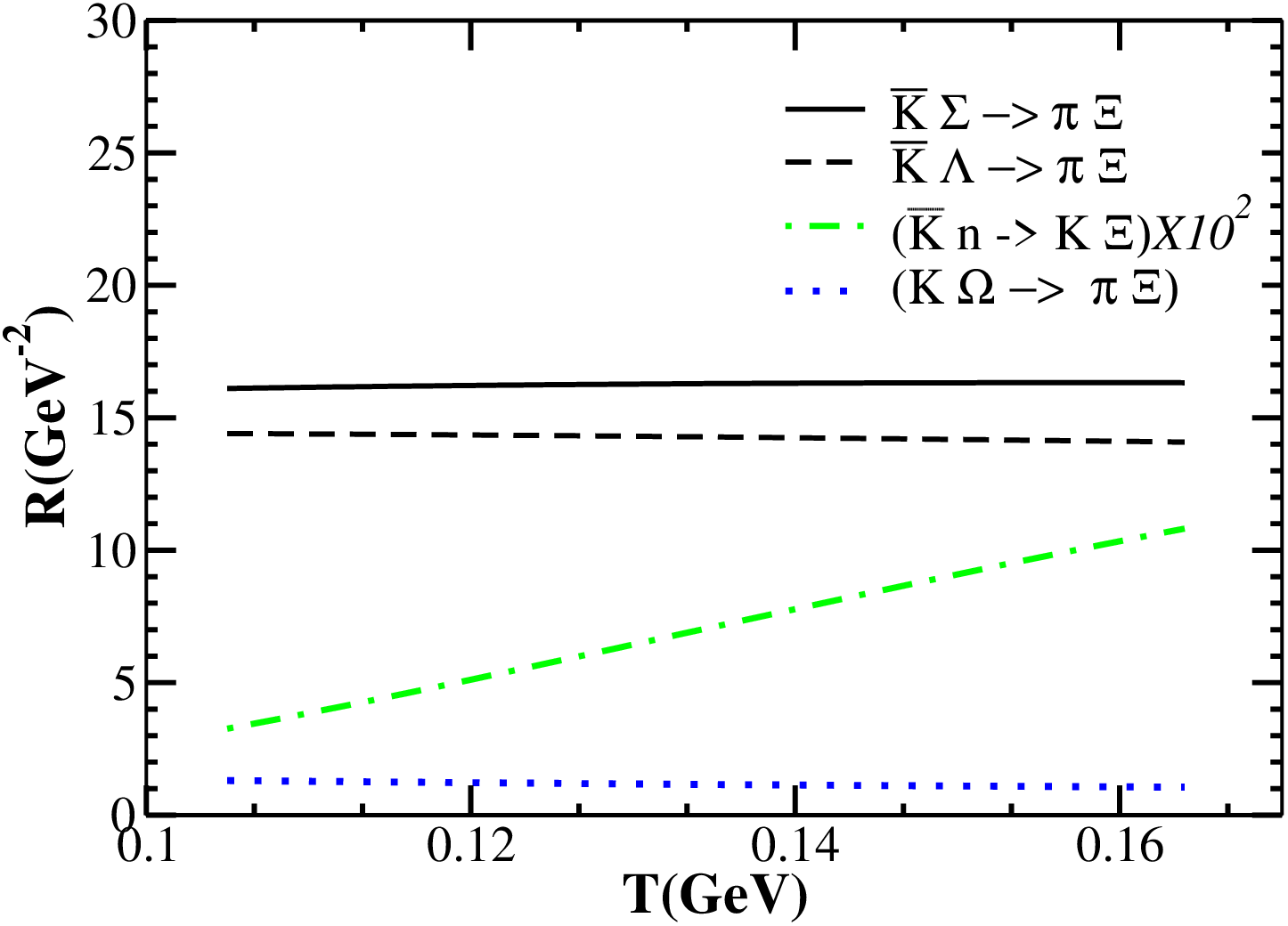}
\caption{Rate, R(T)(=$\langle \sigma v\rangle$)of cascade production with $\bar{K}\Sigma, \bar{K}\Lambda,\bar{K}N, K\Omega$ in the initial channels.}
\label{fig_cascaderate2}
\end{center}
\end{figure}
\begin{figure}[t]
\begin{center}
\includegraphics[scale=0.3]{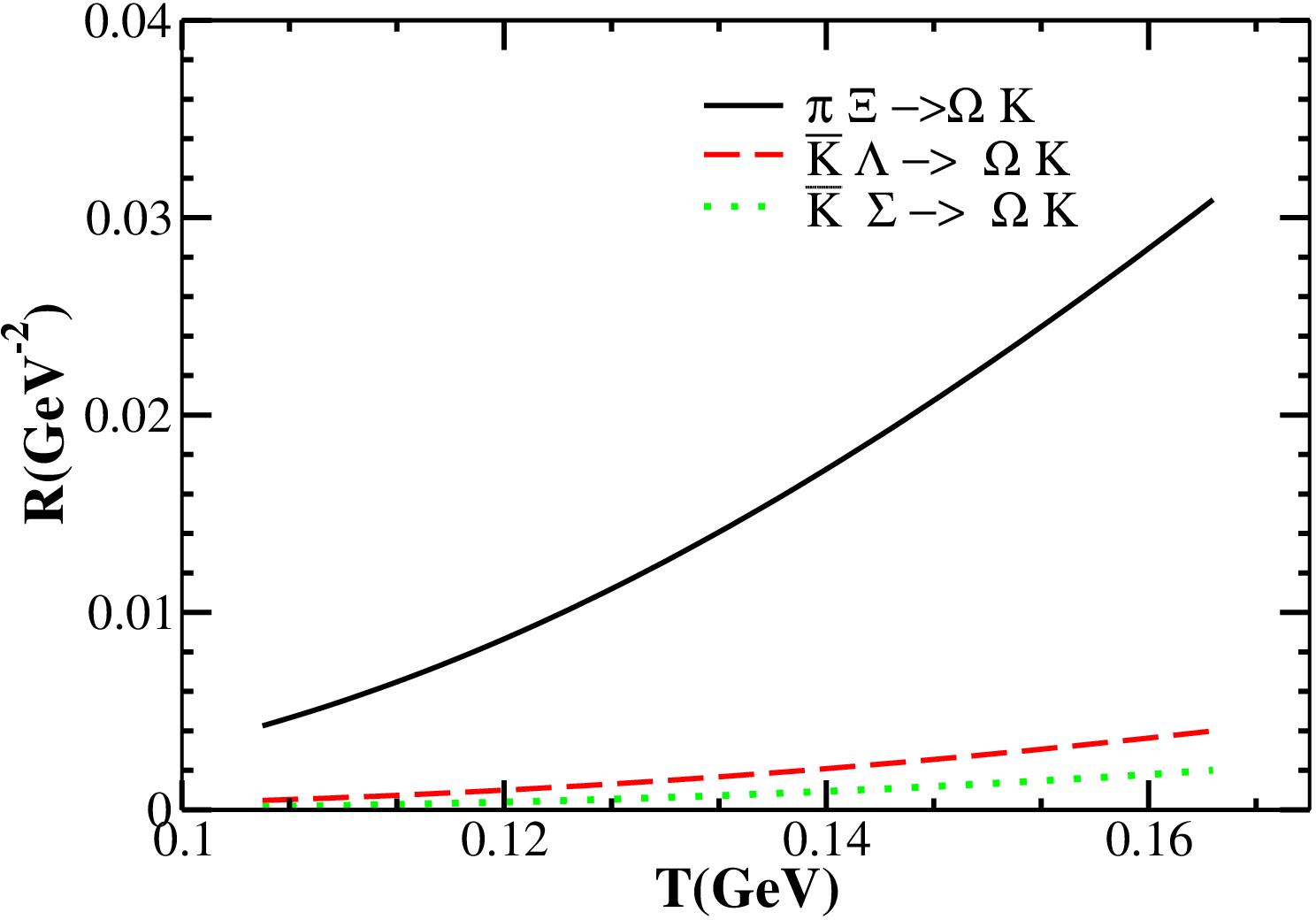}
\caption{Rate (R=$\langle \sigma v\rangle$)of omega production from $\pi\Xi \rightarrow \Omega K$, ${\bar K}\Lambda \rightarrow \Omega K$ and ${\bar K}\Sigma \rightarrow \Omega K$ reactions at various temperature.}
\label{fig_omegarate1}
\end{center}
\end{figure}
\begin{figure}
\begin{center}
\includegraphics[scale=0.3]{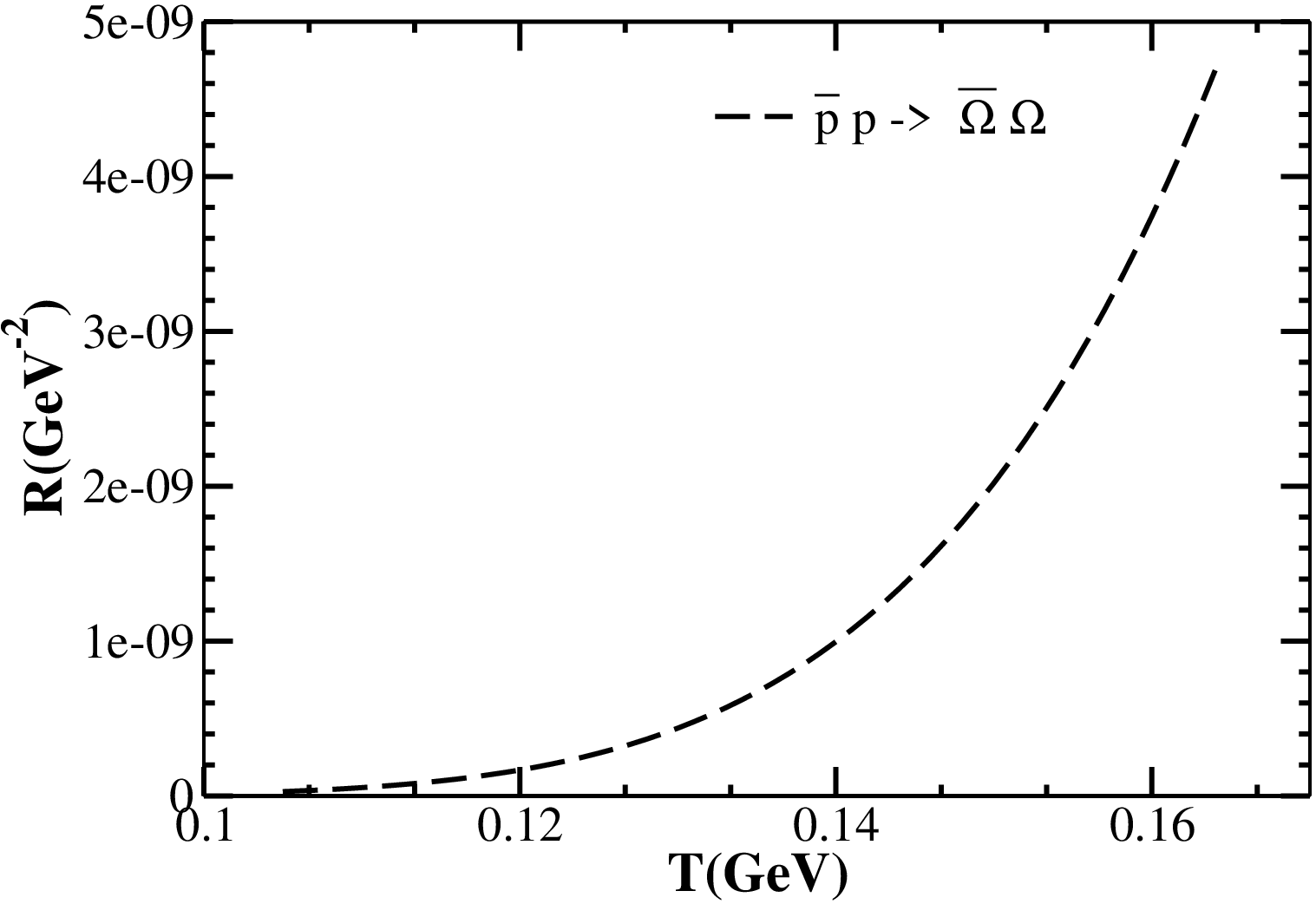}
\caption{Rate of omega production from $p\bar{p} \rightarrow \Omega \bar{\Omega}$  reactions at various temperature.}
\label{fig_omegarate2}
\end{center}
\end{figure}
\subsubsection{\bf Cross sections for $\Omega$ production}
$Omega$ is the hyperon with maximum strangeness and production channels are not understood clearly. However, following channels are considered in this work.
$K^{-} \Lambda  \rightarrow \Omega^{-} K^{0}$, $K^{-}\Sigma^0 \rightarrow \Omega^{-} K^{0}$, 
$\pi^{0}\Xi \rightarrow \Omega^{-}K^{0}$, $p \bar{p} \rightarrow \Omega \bar{\Omega}$.
The reactions like $\Xi Y \rightarrow \Omega N$, $\bar{K} \Xi \rightarrow \Omega \pi$ although produce $\Omega$ but we don't have a clear understanding of the production cross section. However some authors mention that the production is like $\bar{K} N \rightarrow \pi Y$ \cite{koch89} but the necessary experimental coupling is missing. The authors in ~\cite{gaitanos16} mention the $\Omega$ production from $\pi \Xi \rightarrow \Omega K$ ($\pi^0 \Xi^- \rightarrow \Omega^- K^0$), $\bar{K} Y \rightarrow K \Omega$ ($\bar{K} \Lambda \rightarrow K^0 \Omega^{-} $, 
$\bar{K} \Sigma^{0}\rightarrow K^0 \Omega^{-}$)~\cite{gaitanos16} using the cross section from PYTHIA simulation. The cross sections are parametrised as follows,  
\begin{eqnarray}
 \sigma_{K^{-} \Lambda \rightarrow \Omega^{-} K^{0}}&=&a_0+a_1 \, p_{\text{lab}}+a_2\,  p_{\text{lab}}^2+a_3 \, \text{exp}(-a_4 p_{\text{lab}}) \nonumber\\
 \sigma_{K^{-} \Sigma^0 \rightarrow \Omega^{-} K^{0}}&=&b_0+b_1 p_{\text{lab}}+b_2 p_{\text{lab}}^2+b_3 \, \text{exp}(-b_4 p_{\text{lab}}) \nonumber\\
 \sigma_{\pi^{0} \Xi^{-} \rightarrow \Omega^{-} K^{0}}&=&c_0+c_1 \, p_{\text{lab}}+c_2\, p_{\text{lab}}^2+c_3/p_{\text{lab}}+\nonumber\\
 &&c_4/(p_{\text{lab}}^2)+c_5 \text{exp}(-p_{\text{lab}})  
\end{eqnarray}
The parameters $a_i$'s and the validity of crossections in the domain of momentum in laboratory frame($p_{\text{lab}}$) are mentioned in table~\ref{table_parameters}.

The cross section from the annihilation of $p$ and$\bar{p}$ is taken from~\cite{kaidalov94} and reads as follows;
\begin{eqnarray}
 \sigma_{\bar{p} p\rightarrow \Omega^{-} \bar{\Omega}^+}=
 \frac{4^3}{\pi^2}\times\frac{[\sigma_{p \bar{p}\rightarrow \bar{\Lambda}\Lambda}]^3}{\Lambda_1^2}
 \times \exp[\Lambda_1 t_{DO}]
\end{eqnarray}
where $t_{DO}=t_{min}^{\Lambda\Xi}-t_{min}^{p\Lambda}+t_{min}^{\Xi\Omega}-t_{min}^{p\Lambda}$ and $t_{min}^{ij}=-\frac{s}{2}+m_{i}^{2}+m_{j}^{2}+\frac{1}{2}\sqrt{(s-4m_{i}^{2})(s-4m_{j}^{2})}$. For details see \cite{jkn19}. Out of these four channels the proton antiproton producing omega channels is the primary omega producing channel and the rest three channels are secondary channels.
\subsection{\bf Cross sections for other strange hadrons}
The productions of other strange mesons, single strange baryons as follows 
$\pi \pi \rightarrow K \bar{K}$,
$\pi \rho \rightarrow K \bar{K}$,
$\rho \rho \rightarrow K \bar{K}$,
$\pi N \rightarrow \Lambda K$,
$\rho N \rightarrow \Lambda K$,
$\pi N \rightarrow \Sigma K$, 
$\bar{K} N \rightarrow \Lambda \pi$,
$\bar{K} N \rightarrow \Sigma \pi$,
$\bar{p} p \rightarrow \Lambda \bar{\Lambda}$,
$\bar{p} p \rightarrow \Sigma^- \bar{\Sigma^+}$, 
$\bar{p} p \rightarrow K^- \bar{K^+}$,
$N \Xi \rightarrow \Lambda \Lambda$,
$N \Xi \rightarrow \Lambda \Sigma$,
$N \Xi \rightarrow \Sigma \Sigma$,
$K \Xi \rightarrow \bar{K} N$, 
$\pi \Xi \rightarrow \bar{K} \Lambda$,
$\pi \Xi \rightarrow \bar{K} \Sigma$,
$K \Omega \rightarrow K \Sigma$, 
$K\Omega \rightarrow {\bar K}\Lambda$ 
etc. are
considered simultaneously to calculate the multi strange baryons. The cross sections are described in ~\cite{Brown1,amslar08,liprc85,kaidalov94,cugnonnpa84,linpa97,jknprc10}.
 \begin{table}[H]
 \caption{Parameters for $\Omega$ productions}
 \centering
 \footnotesize
  $\sigma_{K^{-} \Lambda^0 \rightarrow \Omega^{-} K^{0}}$ ($1.011\leq P_{lab} (GeV)\leq 6.55$)\\
   \begin{tabular}{|c|c|c|c|c|}
   \hline
   $a_0$ & $a_1$ & $a_2$ & $a_3$ & $a_4$\\
  \hline
    0.155591 & -0.0473326 & 0.00362302 & -0.29776 & 0.917116 \\
  \hline
    \end{tabular}
  $\sigma_{K^{-} \Sigma^0 \rightarrow \Omega^{-} K^{0}}$ ($1.19\leq P_{lab}(GeV)\leq 5.991$)\\
  \begin{tabular}{|c|c|c|c|c|}
  \hline
  $b_0$ & $b_1$ & $b_2$ & $b_3$ & $b_4$\\
  \hline
    0.137027 & -0.0422865 & 0.00327658 & -0.281588 & 0.942457 \\
    \hline
   \end{tabular} 
  $\sigma_{\pi^{0} \Xi \rightarrow \Omega^{-} K^{0}}$ ($1.033\leq P_{lab}(GeV)\leq 5.351$) \\
   \begin{tabular}{|c|c|c|c|c|c| }
   \hline
  $c_0$ & $c_1$ & $c_2$ & $c_3$ & $c_4$ & $c_5$\\
  \hline
  -0.414988 & -0.025499 & 0.00628967 & 2.1816 & -0.639193 & -2.85555 \\
  \hline    
 \end{tabular}
 \label{table_parameters} 
 \end{table}
\subsection{\bf Rate of production}
We consider the thermal rate for the above binary interactions for strangeness production and evolution in hadronic matter. The rate $R$ at a temperature $T$ for a particular channel of reaction of type $a+b \rightarrow c+d$ is given by ~\cite{kapusta86,gondolo91}, 
\begin{eqnarray}
 \langle \sigma v\rangle &=&\frac{T^4}{4m_a^2m_b^2K_2(m_a/T)K_2(m_b/T)}\int _{z_0}^{\infty} \, dz\, [z^2-\nonumber\\
  && ((m_a+m_b)/T)^2][z^2-((m_a-m_b)/T)^2]\sigma K_1(z) \nonumber
\end{eqnarray}
 where $z_0=\text{max}(m_a+m_b,m_c+m_d)/T$, $z=\frac{\sqrt{s}}{T}$, $\sigma$ is the cross section of particular channel of consideration and $m_a, m_b$ are incoming masses with relative velocity $v$(moller). $K$'s are modified bessel functions.
\section{\label{sec:strange_evolve} Strangeness evolution in Hadronic Medium with secondary productions}
The evolution of $\Xi$ and $\Omega$ and their yield in terms of number density are studied using momentum integrated Boltzmann equation or rate equation for a hadronic medium. The equations for all strange hadrons are mentioned below. Each rate equation contains several production terms according to various reaction channels and a dilution term due to the expansion of the system. Pions which contribute maximally to the total entropy of the system provide the thermal background where the strange hadrons are assumed to be away from equilibrium initially. The hadronic system evolves as the temperature falls. The rate equations are as follows. 
\begin{eqnarray}
\frac{dn_{K}}{dt}&=& n_{\pi}n_{\pi}
\langle\sigma v\rangle_{\pi\pi\rightarrow K\bar{K}} 
-n_{K}n_{\bar{K}}\langle\sigma v \rangle_{K\bar{K}\rightarrow \pi\pi} 
\nonumber\\
&& 
+n_{\rho} n_{\rho} \langle\sigma v\rangle_{\rho\rho\rightarrow K\bar{K}} 
-n_{K}n_{\bar{K}}\langle\sigma v\rangle_{K\bar{K}\rightarrow \rho\rho} 
\nonumber\\
&& 
+n_{\pi} n_{\rho}\langle\sigma v\rangle_{\pi\rho\rightarrow K\bar{K}}
-n_{K}n_{\bar{K}}\langle\sigma v\rangle_{K\bar{K}\rightarrow \pi\rho}
\nonumber\\
&& 
+n_{\pi} n_N\langle\sigma v\rangle_{\pi N\rightarrow \Lambda K}
-n_{\Lambda} n_K\langle\sigma v\rangle_{\Lambda K\rightarrow \pi N}
\nonumber\\
&& 
+n_{\rho} n_N\langle\sigma v\rangle_{\rho N\rightarrow \Lambda K}
-n_{\Lambda} n_K\langle\sigma v\rangle_{\Lambda K\rightarrow \rho N}
\nonumber\\
&&
+n_{\pi} n_N\langle\sigma v\rangle_{\pi N\rightarrow \Sigma K}
-n_{\Sigma} n_K\langle\sigma v\rangle_{\Sigma K\rightarrow \pi N}
\nonumber\\
&& 
+n_{\bar K} n_N\langle\sigma v\rangle_{\bar{K}N\rightarrow K\Xi}
 -n_K n_{\Xi}\langle\sigma v\rangle_{K\Xi\rightarrow \bar{K}N}
\nonumber\\
&&
+n_p n_{\bar{p}} \langle\sigma v\rangle_{ p \bar{p}\rightarrow K \bar{K}}
-n_K n_{\bar{K}}\langle\sigma v\rangle_{K \bar{K} \rightarrow p \bar{p}}
\nonumber\\ 
&&
+n_{\bar{K}} n_{\Lambda} \langle\sigma v\rangle_{\bar{K} \Lambda\rightarrow \Omega K}
-n_{\Omega}n_{K}\langle\sigma v\rangle_{\Omega K \rightarrow \bar{K}\Lambda}
\nonumber\\
&&
+n_{\bar{K}} n_{\Sigma} \langle\sigma v\rangle_{\bar{K} \Sigma\rightarrow \Omega K}
-n_{\Omega}n_{K}\langle\sigma v\rangle_{\Omega K \rightarrow \bar{K}\Sigma}
\nonumber\\
&&
+n_{\pi} n_{\Xi}\langle\sigma v\rangle_{\pi \Xi \rightarrow K\Omega}
-n_{\Omega} n_{K}\langle\sigma v\rangle_{ \Omega K \rightarrow \pi \Xi}
-\frac{n_K}{t} \nonumber
\end{eqnarray}
\begin{eqnarray}
\frac{dn_{\bar{K}}}{dt}&=& n_{\pi}n_{\pi}
\langle\sigma v\rangle_{\pi\pi\rightarrow K\bar{K}} 
-n_{K}n_{\bar{K}}\langle\sigma v \rangle_{K\bar{K}\rightarrow \pi\pi}
\nonumber\\
&&
+n_{\rho} n_{\rho} \langle\sigma v\rangle_{\rho\rho\rightarrow K\bar{K}} 
-n_{K}n_{\bar{K}}\langle\sigma v\rangle_{K\bar{K}\rightarrow \rho\rho} 
\nonumber\\
&&
+n_{\pi} n_{\rho}\langle\sigma v\rangle_{\pi\rho\rightarrow K\bar{K}}
-n_{K}n_{\bar{K}}\langle\sigma v\rangle_{K\bar{K}\rightarrow \pi\rho}
\nonumber\\
&&
-n_{\bar{K}} n_N \langle\sigma v\rangle_{\bar{K}N\rightarrow \Lambda \pi}
+n_{\Lambda}n_{\pi}\langle\sigma v\rangle_{\Lambda \pi\rightarrow \bar{K}N}
\nonumber\\
&&
-n_{\bar{K}} n_N \langle\sigma v\rangle_{\bar{K}N\rightarrow \Sigma\pi}
+n_{\Sigma}n_{\pi}\langle\sigma v\rangle_{\Sigma \pi\rightarrow \bar{K}N}
\nonumber\\
&&
-n_{\bar K} n_N\langle\sigma v\rangle_{\bar{K}N\rightarrow K\Xi}
+n_K n_{\Xi}\langle\sigma v\rangle_{K\Xi\rightarrow \bar{K}N}
\nonumber\\
&& 
-n_{\bar{K}} n_{\Lambda}\langle\sigma v\rangle_{\bar{K}\Lambda\rightarrow \pi\Xi}
+n_{\pi} n_{\Xi}\langle\sigma v\rangle_{\pi\Xi \rightarrow \bar{K}\Lambda}
\nonumber\\
&& 
-n_{\bar{K}} n_{\Sigma}\langle\sigma v\rangle_{\bar{K}\Sigma\rightarrow \pi\Xi}
+n_{\pi} n_{\Xi}\langle\sigma v\rangle_{\pi\Xi \rightarrow \bar{K}\Sigma}
\nonumber\\
&&
+n_p n_{\bar{p}} \langle\sigma v\rangle_{ p\bar{p} \rightarrow  K \bar{K}}
-n_{K}n_{\bar{K}}\langle\sigma v\rangle_{ K \bar{K} \rightarrow p \bar{p}}
\nonumber\\ 
&&
-n_{\bar{K}} n_{\Lambda} \langle\sigma v\rangle_{\bar{K} \Lambda\rightarrow \Omega K}
+n_{\Omega}n_{K}\langle\sigma v\rangle_{\Omega K \rightarrow \bar{K}\Lambda}
\nonumber\\
&&
-n_{\bar{K}} n_{\Sigma} \langle\sigma v\rangle_{\bar{K} \Sigma\rightarrow \Omega K}
+n_{\Omega}n_{K}\langle\sigma v\rangle_{\Omega K \rightarrow \bar{K}\Sigma}
-\frac{n_{\bar{K}}}{t} \nonumber
\end{eqnarray}
\begin{eqnarray}
 \frac{dn_{\Lambda}}{dt}&=&
n_{\pi} n_N\langle\sigma v\rangle_{\pi N\rightarrow \Lambda K}
-n_{\Lambda} n_K\langle\sigma v\rangle_{\Lambda K\rightarrow \pi N}
\nonumber\\
&& 
+n_{\rho} n_N\langle\sigma v\rangle_{\rho N\rightarrow \Lambda K}
-n_{\Lambda} n_K\langle\sigma v\rangle_{\Lambda K\rightarrow \rho N}
\nonumber\\
&&
-n_{\Lambda} n_{\Lambda}\langle\sigma v\rangle_{\Lambda\Lambda\rightarrow N\Xi}
+n_N n_{\Xi}\langle\sigma v\rangle_{N \Xi\rightarrow \Lambda\Lambda}\nonumber\\
&&-n_{\Lambda} n_{\Sigma}\langle\sigma v\rangle_{\Lambda\Sigma\rightarrow N\Xi}
+n_N n_{\Xi}\langle\sigma v\rangle_{N \Xi\rightarrow \Lambda\Sigma}
\nonumber\\
&&
-n_{\bar K} n_{\Lambda}\langle\sigma v\rangle_{\bar{K}\Lambda\rightarrow \pi\Xi}
+n_{\pi} n_{\Xi}\langle\sigma v\rangle_{\pi\Xi \rightarrow \bar{K}\Lambda}
\nonumber\\
&&
+n_{\bar K} n_{N}\langle\sigma v\rangle_{\bar{K}N\rightarrow {\Lambda}\pi}
-n_{\Lambda} n_{\pi}\langle\sigma v\rangle_{\Lambda\pi \rightarrow \bar{K}N}
\nonumber\\
&&
+n_p n_{\bar{p}}\langle\sigma v\rangle_{p \bar{p}\rightarrow \Lambda\bar{\Lambda}}
 -n_{\Lambda} n_{\bar{\Lambda}}\langle\sigma v\rangle_{ \Lambda \bar{\Lambda}\rightarrow p\bar{p}} 
\nonumber\\
&&+n_{K} n_{\Omega}\langle\sigma v\rangle_{K{\Omega} \rightarrow {\bar K}{\Lambda}}
 -n_{{\bar K}} n_{\Lambda}\langle\sigma v\rangle_{{\bar K} {\Lambda}\rightarrow K \Omega}-
 \frac{n_{\Lambda}}{t} \nonumber
\end{eqnarray}
\begin{eqnarray}
 \frac{dn_{\Sigma}}{dt}&=&
n_{\pi} n_N\langle\sigma v\rangle_{\pi N\rightarrow \Sigma K}
-n_{\Sigma} n_K\langle\sigma v\rangle_{\Sigma K\rightarrow \pi N}
\nonumber\\
&&
-n_{\Lambda} n_{\Sigma}\langle\sigma v\rangle_{\Lambda\Sigma\rightarrow N\Xi}
+n_N n_{\Xi}\langle\sigma v\rangle_{N \Xi\rightarrow \Lambda\Sigma}
\nonumber\\
&&
-n_{\Sigma} n_{\Sigma}\langle\sigma v\rangle_{\Sigma\Sigma\rightarrow N\Xi}
+n_N n_{\Xi}\langle\sigma v\rangle_{N \Xi\rightarrow \Sigma\Sigma}\nonumber\\
&&
-n_{\bar{K}} n_{\Sigma}\langle\sigma v\rangle_{\bar{K}\Sigma\rightarrow \pi\Xi}
 +n_{\pi} n_{\Xi}\langle\sigma v\rangle_{\pi\Xi \rightarrow \bar{K}\Sigma} 
\nonumber\\
&&
+n_{\bar K} n_{N}\langle\sigma v\rangle_{\bar{K}N\rightarrow {\Sigma}\pi}
-n_{\Sigma} n_{\pi}\langle\sigma v\rangle_{\Sigma\pi \rightarrow \bar{K}N}
\nonumber\\
&&
 +n_p n_{\bar{p}}\langle\sigma v\rangle_{p \bar{p}\rightarrow \Sigma\bar{\Sigma}}
 -n_{\Sigma} n_{\bar{\Sigma}}\langle\sigma v\rangle_{ \Sigma \bar{\Sigma}\rightarrow p\bar{p}}
\nonumber\\
&&
 +n_{K} n_{\Omega}\langle\sigma v\rangle_{K{\Omega} \rightarrow \bar{K}{\Sigma}}
 -n_{\bar{K}} n_{\Sigma}\langle\sigma v\rangle_{{\bar K} {\Sigma}\rightarrow K \Omega}-
 \frac{n_{\Sigma}}{t} \nonumber
\end{eqnarray}
\begin{eqnarray}
\frac{dn_{\Xi}}{dt}&=&
n_{\Lambda} n_{\Lambda}\langle\sigma v\rangle_{\Lambda\Lambda\rightarrow N\Xi}
-n_N n_{\Xi}\langle\sigma v\rangle_{N \Xi\rightarrow \Lambda\Lambda}
\nonumber\\
&&
+n_{\Lambda} n_{\Sigma}\langle\sigma v\rangle_{\Lambda\Sigma\rightarrow N\Xi}
-n_N n_{\Xi}\langle\sigma v\rangle_{N \Xi\rightarrow \Lambda\Sigma}
\nonumber\\
&&
+n_{\Sigma} n_{\Sigma}\langle\sigma v\rangle_{\Sigma\Sigma\rightarrow N\Xi}
-n_N n_{\Xi}\langle\sigma v\rangle_{N \Xi\rightarrow \Sigma\Sigma}\nonumber\\
&&
 +n_{\bar K} n_N\langle\sigma v\rangle_{\bar{K}N\rightarrow K\Xi}
 -n_K n_{\Xi}\langle\sigma v\rangle_{K\Xi\rightarrow \bar{K}N}
\nonumber\\
&& 
+n_{\bar{K}} n_{\Lambda}\langle\sigma v\rangle_{\bar{K}\Lambda\rightarrow \pi\Xi}
 -n_{\pi} n_{\Xi}\langle\sigma v\rangle_{\pi\Xi \rightarrow \bar{K}\Lambda}
\nonumber\\
&& 
+n_{\bar{K}} n_{\Sigma}\langle\sigma v\rangle_{\bar{K}\Sigma\rightarrow \pi\Xi}
-n_{\pi} n_{\Xi}\langle\sigma v\rangle_{\pi\Xi \rightarrow \bar{K}\Sigma}\nonumber\\
&&+n_p n_{\bar{p}}\langle\sigma v\rangle_{p \bar{p}\rightarrow \Xi\bar{\Xi}}
 -n_{\Xi} n_{\bar{\Xi}}\langle\sigma v\rangle_{ \Xi \bar{\Xi}\rightarrow p\bar{p}} 
\nonumber\\
&&
 +n_{\Omega} n_{K}\langle\sigma v\rangle_{ \Omega K\rightarrow {\pi}{\Xi}}-
 n_{\pi} n_{\Xi}\langle\sigma v\rangle_{\pi {\Xi}\rightarrow \Omega K}-
 \frac{n_{\Xi}}{t} \nonumber
\end{eqnarray}
\begin{eqnarray}
\frac{dn_{\Omega}}{dt}&=&
n_p n_{\bar{p}}\langle\sigma v\rangle_{p \bar{p}\rightarrow \Omega\bar{\Omega}}
 -n_{\Omega} n_{\bar{\Omega}}\langle\sigma v\rangle_{ \Omega \bar{\Omega}\rightarrow p\bar{p}}
\nonumber\\
&&
 +n_{\pi} n_{\Xi}\langle\sigma v\rangle_{\pi {\Xi}\rightarrow \Omega K}
 -n_{\Omega} n_{K}\langle\sigma v\rangle_{ \Omega K\rightarrow {\pi}{\Xi}} 
\nonumber\\
&&
 +n_{{\bar K}} n_{\Lambda}\langle\sigma v\rangle_{{\bar K} {\Lambda}\rightarrow K \Omega}
 -n_{K} n_{\Omega}\langle\sigma v\rangle_{ K\Omega\rightarrow{\bar K}{\Lambda}}\nonumber\\
 &&+n_{\bar K} n_{\Sigma}\langle\sigma v\rangle_{{\bar K} {\Sigma}\rightarrow K \Omega}
 -n_{K} n_{\Omega}\langle\sigma v\rangle_{K\Omega\rightarrow{\bar K}{\Sigma}} - \frac{n_{\Omega}}{t}
 \end{eqnarray}
Along with the rate equations the evolution of baryonic chemical potential and temperature have also been considered with relativistic bjorken hydrodynamic expansion\cite{bjorken} and the chemical potential has been constrained with the values obtained from statistical hadronization model. 
 \begin{table}[t]
 \caption{initial conditions (Freeze out temperatures, $T_F$) for various multiplicities for various scenarios-I, II, III, IV, V}
 \centering 
 \footnotesize
 \begin{tabular}{ |c|c|c|c|c|c|c| } 
  \hline
  $dn_{ch}/d\eta$ & $N_{part}$ & I & II & III & IV & V \\
   & & $T_{f_{1}}$ & $ T_{f_{2}}$ & $T_{f_{3}}$ & $T_{f_{4}}$ & $T_{f_{5}}$  \\
  & &$\Xi,\Omega$&$\Xi,\Omega$&$\Xi,\Omega$&$\Xi,\Omega$& $\Xi,\Omega$  \\ 
  \hline
  1447.5  & 356.1 &    0.144   &     0.144   &    0.144   &    0.154  &    0.134,~0.145 \\
  \hline
  966     & 260.1 &    0.142   &     0.144   &    0.144   &    0.154   &    0.141,~0.144  \\
  \hline
  537.5   & 157.2 &    0.140   &     0.144   &    0.144   &    0.154   &    0.143,~0.143  \\
  \hline
  205     & 68.6 &    0.132   &     0.144  &    0.144   &    0.154   &    0.137,~0.137   \\
  \hline
  55      & 22.5 &    0.116   &     0.144   &    0.144   &    0.154   &    0.118,~0.118  \\
  \hline
 \end{tabular}
 \label{table_IC}
 \end{table}
\section{\label{sec:results} Results}
The rate, $R$(=$\langle \sigma v\rangle$) of multi-strange hadron productions have been evaluated for all mentioned channels considering the cross sections mentioned in earlier section. The cascade production rates are displayed in Figs.\ref{fig_cascaderate1} \& \ref{fig_cascaderate2}. The rates from $\Lambda\Lambda \rightarrow N\Xi$, $\Lambda\Sigma \rightarrow N\Xi$, $\Sigma\Sigma \rightarrow N\Xi$ do not vary much with temperature. The cross sections for these reactions decrease very slowly with the centre of mass energy of the colliding channel ( Eq.~\ref{casccross2a}) beyond the threshold, while the centre of mass energy increases slowly with the temperature within the range where thermal rates have been shown. Hence the rates for these reactions appear to be constant (although slightly increase with temperature) when the Boltzmann factor is considered. However, it is found that contributions from $\Lambda \Sigma$ interactions is 7-8 times larger than $\Sigma \Sigma$ and 2-3 times larger than $\Lambda \Lambda$.

The rate from $\bar{K} \Sigma \rightarrow \pi\Xi$ is found to be more than $\bar{K} \Lambda \rightarrow \pi\Xi$ and $\bar{K} N \rightarrow K\Xi$ as shown in the plots displayed in Fig.\ref{fig_cascaderate2}. The rates from $\bar{K} \Sigma$ and $\bar{K} N$ are also found not to vary much within this temperature range. The production from $K\Omega \rightarrow \pi \Xi$ does not contribute much as shown in Fig.~\ref{fig_cascaderate2}. $\Lambda \Sigma \rightarrow N\Xi$ is the dominant channel for cascade productions and the net cascade yield is decided by $\Lambda, \Sigma$ and $K$ interactions. The rate of $\Xi$ and $\Omega$ productions from non strange hadrons as initial channels are less compared to strangeness exchange reactions as their cross sections are less. One can have the information from the comparison of the production from the channels $p p \rightarrow \Xi \bar{\Xi}$ and $\Lambda \Lambda \rightarrow N \Xi$ or $\bar{K} \Lambda \rightarrow \pi \Xi$. The rate of production in case of $p p \rightarrow \Xi \bar{\Xi}$ is $10^6$ times less. 

The omega productions from $\pi\Xi\rightarrow K\Omega$, ${\bar K}\Lambda\rightarrow K\Omega$, ${\bar K}\Sigma\rightarrow K\Omega$ and $p\bar{p}\rightarrow \Omega\bar{\Omega}$ are shown in Figs.~\ref{fig_omegarate1} \&\ref{fig_omegarate2}. The contribution of $\pi \Xi \rightarrow K \Omega$ is the dominant one as the threshold is less compared to other channels and the pion abundance is more.  

Cascade and omega yields have been calculated from momentum integrated Boltzmann transport equations which considers the production and evolution of all strange hadrons simultaneously. Yield of these particles are normalized with thermal pions. The study has been done for various initial conditions. The initial number densities of strange hadrons are unknown parameters and considered to be away from equilibrium value initially. A hadronic system is assumed to be started at $T_c$=155 MeV. The value taken from the recent first principle calculation of quantum chromodynaimcs based on lattice computation~\cite{swagato17}, which suggests a value of 154$\pm$9 MeV as the transition temperature. Then different scenarios are assumed with different initial conditions to analyse the data obtained from Pb+Pb collisions at $\sqrt{s_{NN}}$=2.76 TeV, LHC energy ~\cite{multistrange_alice_plb14,alicenature17}. Initial time is constrained with multiplicity. The yields of cascade and omega baryons have been measured for various multiplicities and normalised with charged pion data. The corresponding centralities and $N_{\text{part}}$ for various multiplicities are shown in table~\ref{table_IC}. Theoretical results are obtained for the following scenarios.

\begin{figure}
\begin{center}
\includegraphics[scale=0.3]{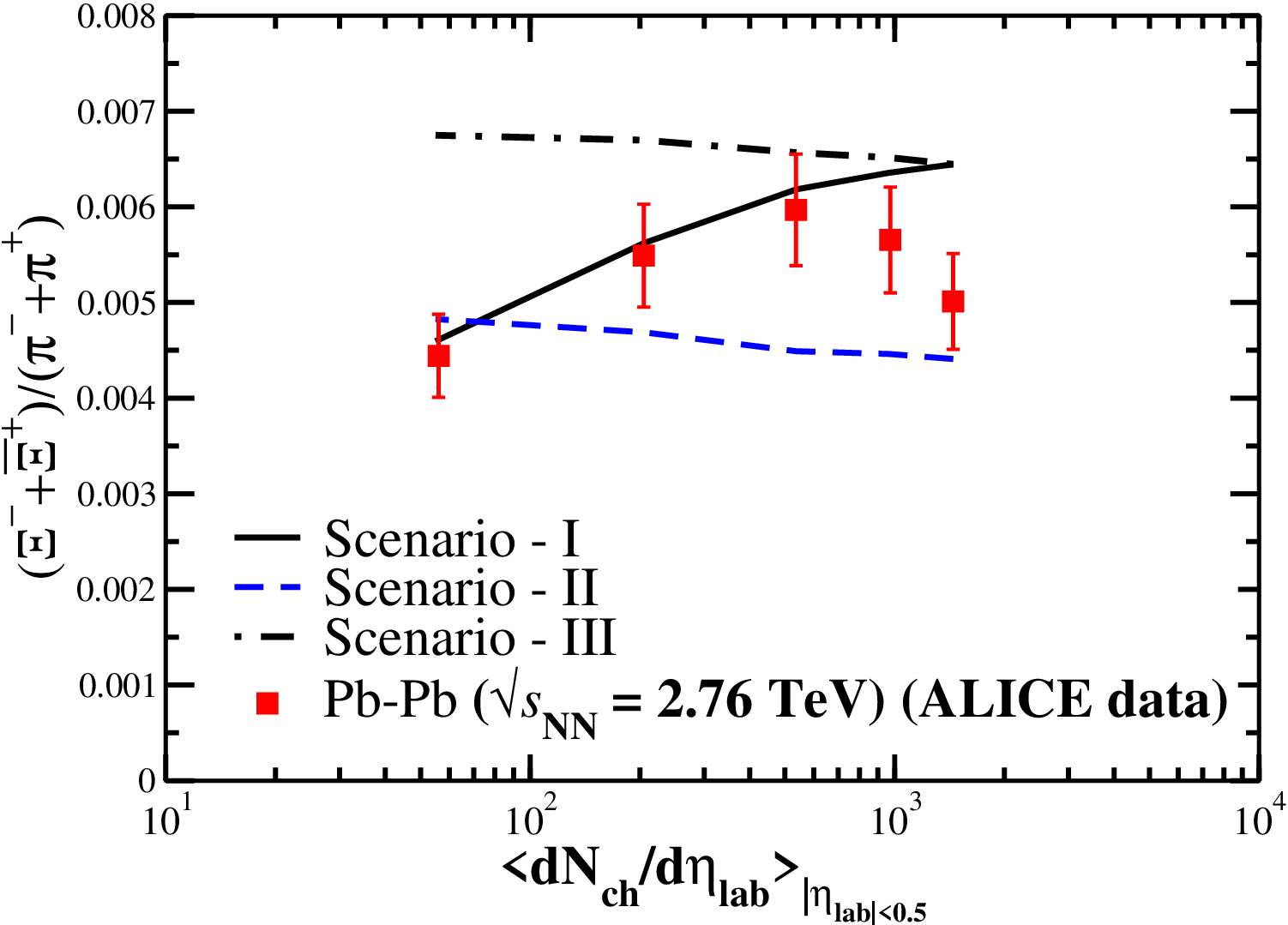}
\caption{Ratio of the yield of cascade to pion with multiplicity (centrality). The solid points are data points from 2.76 TeV Pb+Pb collisions measured by ALICE collaboration. The solid lines are the results of theoretical calculation with initial condition for scenario-I, II and III.}
\label{fig_multistrangebypiratio-1}
\end{center}
\end{figure}

\begin{figure}[t]
\begin{center}
\includegraphics[scale=0.3]{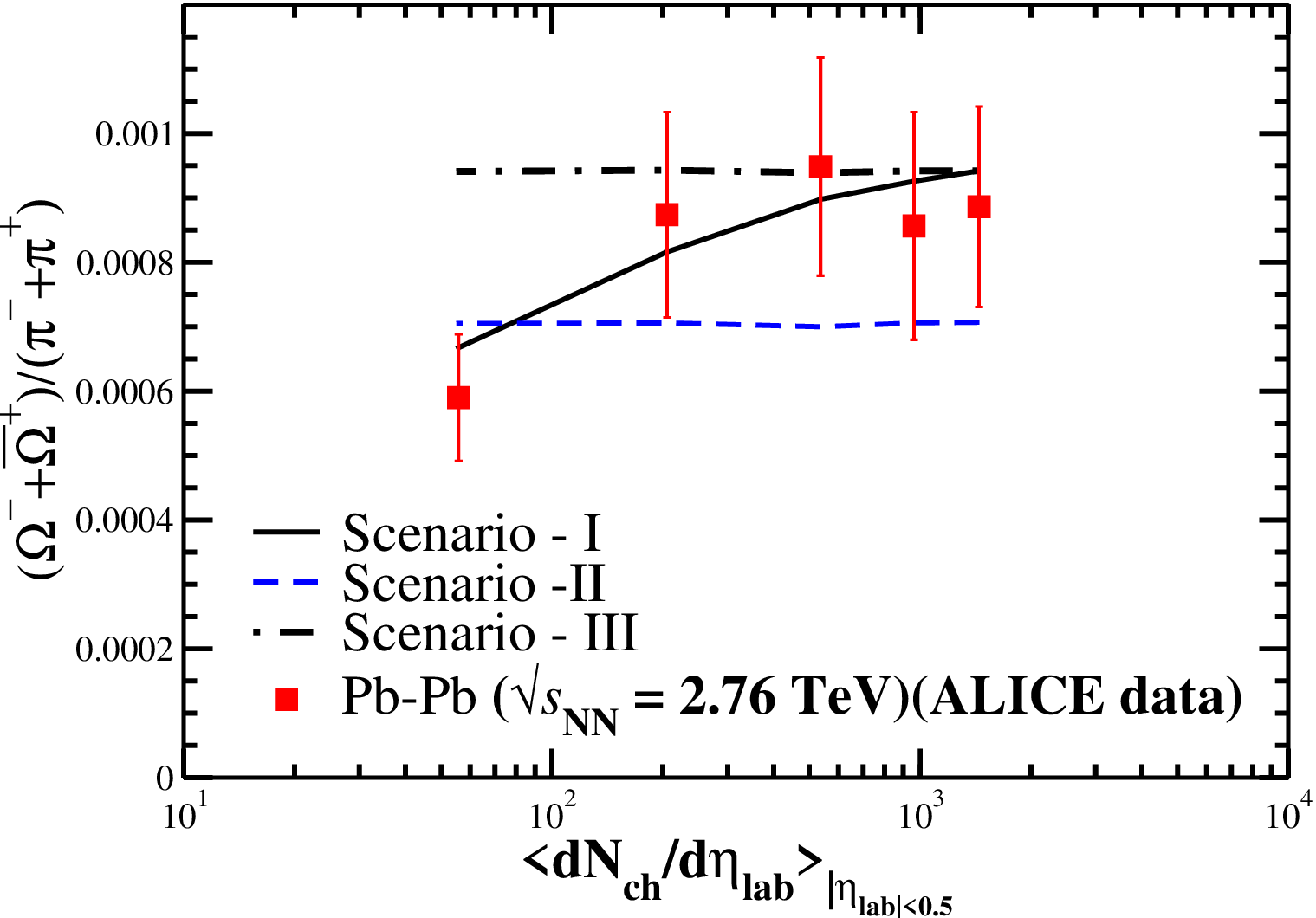}
\caption{Ratio of the yield of omega to pion with multiplicity (centrality). The solid points are data points from 2.76 TeV Pb+Pb collisions measured by ALICE collaboration. The solid lines are the results of theoretical calculation with initial condition for scenario-I, II and III.}
\label{fig_multistrangebypiratio-2}
\end{center}
\end{figure}

In scenario-I, initial number density is assumed to be 20 $\%$ away from equilibrium. Various freeze out temperatures ($T_F$) have been considered for various multiplicities with velocity of sound $c_s^2$=1/5. The results in terms of the ratio of the yield of $(\Xi^- +\bar{\Xi^+})$ and $(\Omega^-+\bar{\Omega^+})$ to $(\pi^++\pi^-)$ are shown in Figs.~\ref{fig_multistrangebypiratio-1} \& \ref{fig_multistrangebypiratio-2}. The filled symbols are data measured by ALICE collaboration and taken from ~\cite{alicenature17,multistrange_alice_plb14,gyula_alice} and the solid line is the result of theoretical calculation. Higher freeze out temperature is considered for higher multiplicity in scenario-I and the values have been tabulated in table~\ref{table_IC}. The ratios of $(\Omega^-+\overline{\Omega^+})/({\pi^++\pi^-})$ are explained quite successfully for all multiplicities with this initial condition. However, the ratio of     
$(\Xi^-+\overline{\Xi^+})/({\pi^++\pi^-})$ falis to explain the top two data points with higher multiplicities.

Similarly, in Scenario-II, the system is allowed to evolve with initial cascade and omega number densities, 40$\%$ away from the equilibrium value. Here the results are analysed with a constant freeze out temperature $T_F$=144 MeV for all multiplicities. But the evaluation does not reproduce the data. Upon varying the initial number densities to be 20 $\%$ away from equilibrium value but with constant $T_F$=144 MeV for all multiplicities, it doesn't explain the data which is depicted as scenario-III in Fig.\ref{fig_multistrangebypiratio-1}. But it gives a clue to look for a constant freezeout scenario.  

\begin{figure}
\begin{center}
\includegraphics[scale=0.3]{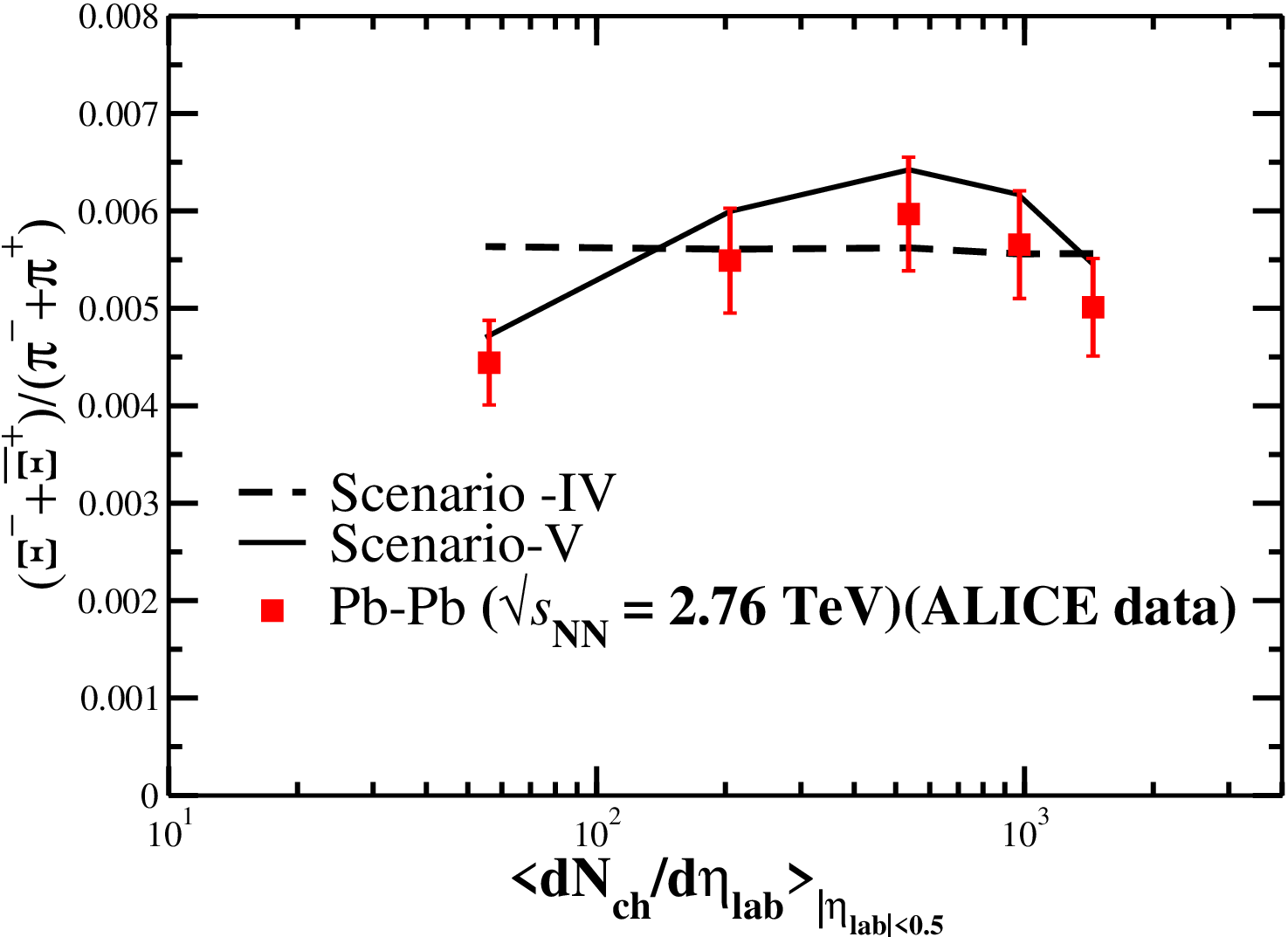}
\caption{Ratio of the yield of cascade to pion with multiplicity (centrality). The solid lines are the results of theoretical calculation with initial condition for scenario-IV \& V.}
\label{fig_multistrangebypiratio-3}
\end{center}
\end{figure}

\begin{figure}[t]
\begin{center}
\includegraphics[scale=0.3]{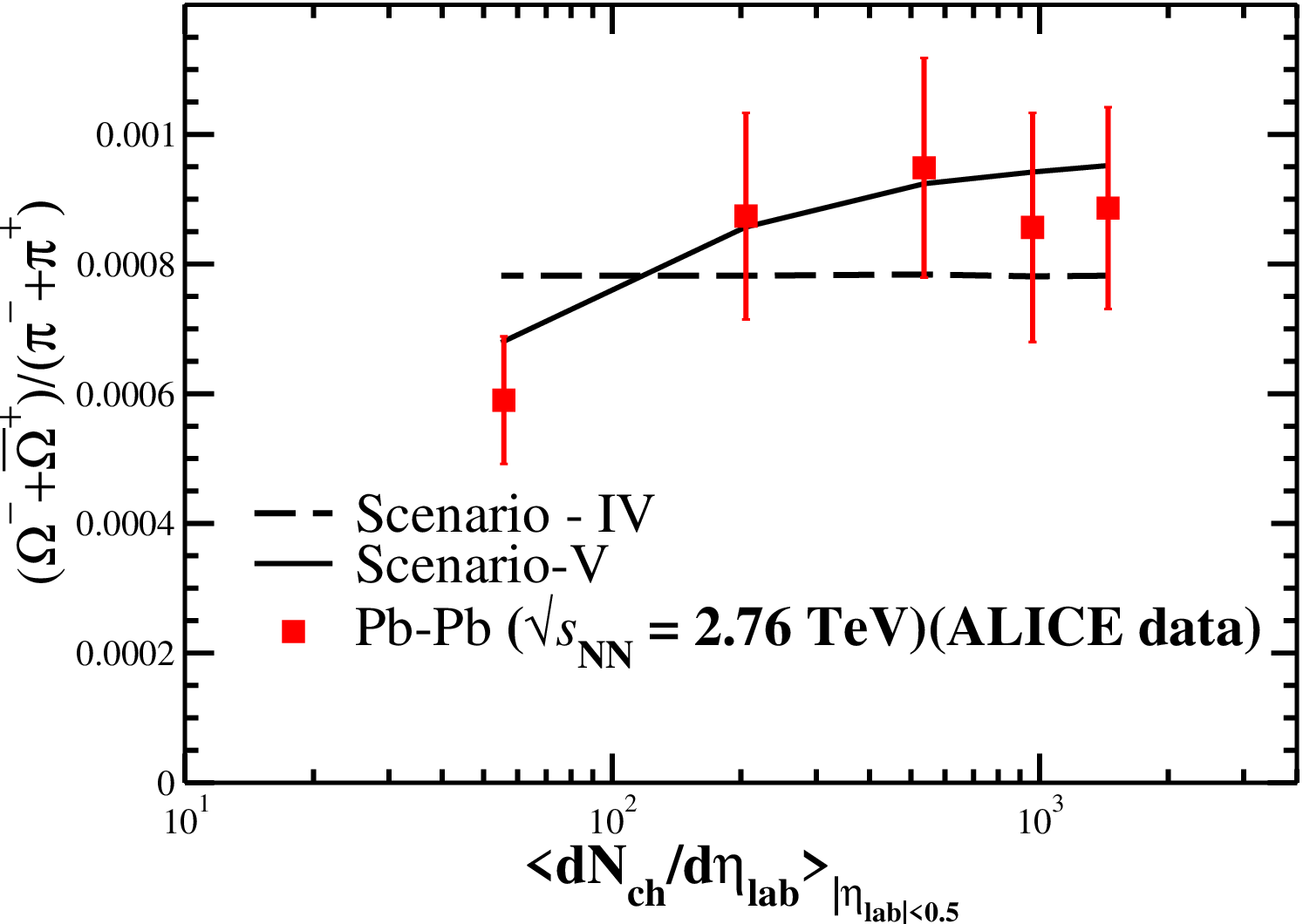}
\caption{Ratio of the yield of omega to pion with multiplicity (centrality). The solid lines are the results of theoretical calculation with initial condition for scenario-IV \& V.}
\label{fig_multistrangebypiratio-4}
\end{center}
\end{figure}

The results have also been obtained for constant $T_F$=154 MeV as there is a thermal model prediction, which has been pointed out in the article(Fig.4 of the article) by Adam {\it et al.} for ALICE collaboration ~\cite{alice_adam2016}. In this case, we assume strange initial densities to be 40$\%$ away from the equilibrium values. This is depicted as Scenario-IV in Figs.\ref{fig_multistrangebypiratio-3} \& \ref{fig_multistrangebypiratio-4}. Here the yield ratio of $\Xi, \Omega$ to $\pi$ remains almost constant with multiplicity and explains the data points excluding the measurement for the lowest multiplicity. The constant freeze out scenario is then ruled out. 
This constant freeze out scenario almost generates similar value of the yield ratio for various multiplicities. That is because the rates of production for cascade and omega do not change much with temperature. It may also tell that the final state effect is dominant for the yield at such energies. Finally, scenario-V considers the initial condition with different freeze out temperature which explains the data nicely with initial number densities 20$\%$ away from equilibrium values and plotted in Fig.\ref{fig_multistrangebypiratio-4}.

The velocity of sound given by $c_s^2$=1/5 is consiedred for the above calculations. Considering $c_s^2$=1/3 through out the evolution, the yields for cascade and omega have also been calculated with initial number densities 20 $\%$ and 40 $\%$ away from equilibrium. However the theoretical estimation overestimates the data for all multiplicities. Hence $c_s^2$=1/3 has been ruled out for hadronic phase here. The yield of $\Xi$ and $\Omega$ depends very strongly on the equation of state or velocity of sound. Fast equation of state or high value of velocity of sound leads to overproduction in the system with present evolution. Hence the calculation over estimates $\Xi/\pi$ or $\Omega/\pi$ data.
\section{\label{sec:summary} Summary}
The $\Xi$ \& $\Omega$ productions have been evaluated microscopically for the first time for Pb-Pb collisions at $\sqrt{s_{NN}}$=2.76 TeV using rate equation considering cross sections from various possible hadronic interactions. The details of the cross sections are discussed referring available literatures, where most of them are constrained with experimental observations. The thermal rates for multi strange hadrons are shown and the yields have been calculated for various initial conditions using rate equations. Largely, the conditions that explain the data of omega suggests a lower freeze out temperature at smaller multiplicity(scenario-I \& V) and it increases with $dN_{ch}/d\eta$, that is when one moves from peripheral to central collisions or from a region of larger overlap to a region of smaller overlap of the colliding nuclei. In case of $\Xi$ the deviation happens at large multiplicities, the explanation requires a lower freeze out (scenario-V). Calculation with a constant freeze out temperature $T_F$ =154 MeV and initial densities 20\% away from the equilibrium values (scenario-IV) also explains most of the data points putting a question mark on the similarity of systems produced in different colliding energies with same multiplicity. This motivates to go for an investigation for small systems with similar multiplicities.

It has been observed that, the calculation with $c_s^2$=1/3 fails to reproduce the data, which basically overestimates for all multiplicities. The yield depends strongly on velocity of sound. Fast equation of state leads to a over production of multistrange hadrons. Incorporation of $c_s^2(T)$ may improve the calculation. When the calculation is extended to analyse the yield of other hadrons $K, \Lambda$, a multiple freeze-out scenario is emerged for Pb-Pb collisions at LHC energy\cite{jkn19}. 

{\it This article with microscopic calculation will provide a guideline to the key question that whether the systems produced in different colliding energies with similar multiplicty are similar are not, which may be addressed in future work}. 
 
{\bf Acknowledgment:} 
Author P. Ghosh thanks VECC for partial support from CNT project vide no. 3/5/2012/VECC/R\&D-I/14802 during the stay at VECC.

\end{document}